\documentclass{ws-ijmpcs}
\usepackage{url}

\def\trimmarks{}

\begin{document}

\markboth{J.W. Qiu}
{Electron-Ion Collider}

\catchline{}{}{}{}{}
%
%%%%%%%%%%%%%%%%%%%%%%%%%%%%%%%%%%%%%%%%%%%%%%%%%%%%%

\title{Electron-Ion Collider - taking us to the next QCD frontier
}

\author{Jian-Wei Qiu
}

\address{Physics Department, 
		Brookhaven National Laboratory,
		Upton, NY 11973-5000, U.S.A.
\\
and\\
C.N.\ Yang Institute for Theoretical Physics
		and Department of Physics and Astronomy,\\
             	Stony Brook University,
             	Stony Brook, NY 11794-3840, USA
\\
jqiu@bnl.gov}

\maketitle

\begin{history}
\received{November 27, 2014}
\revised{Day Month Year}
\published{Day Month Year}
\end{history}

\begin{abstract}
In this talk, I demonstrate that the proposed Electron-Ion Collider (EIC) 
will be an ideal and unique future facility to address many overarching questions 
about QCD and strong interaction physics at one place.  
The EIC will be the world's first polarized electron-proton (and light ion), 
as well as the first electron-nucleus collider at flexible collision energies.  
With its high luminosity and beam polarization, 
the EIC distinguishes itself from HERA and the other fixed target electron-hadron facilities 
around the world.  The EIC is capable of taking us to the next QCD frontier 
to explore the glue that binds us all.
\keywords{Electron-Ion Collider, QCD, hadron structure, gluon saturation.}
\end{abstract}

\ccode{PACS numbers:}

%%%%%%%%%%%%%%%%%%%%%%%%%%%%%%%%%%%%%%%%%%%%%%%%%%%%%
% INTRODUCTION
\section{Introduction}
\label{sec:intro}	

Understanding the evolution of our universe from its origin (the Big Bang) to 
its current state, and the fundamental structure of all matters within it, 
is one of the most central goals of human scientific endeavor.  
The evolution is intimately tied to the properties of and the dynamics between 
fundamental particles that existed in each of its phases. 
The structure of matter encodes all the history and secrets of the evolution 
waiting for us to explore.  

It is the discovery of the atomic structure and the nucleus by the Rutherford experiment 
over 100 years ago that led to the discovery of quantum mechanics and 
the quantum world, which forever changed our view of the universe that we are living in.  
It is the modern ``Rutherford'' experiment, scattering between an electron and a proton, 
performed at Stanford Linear Accelerator Center (SLAC) in the sixties  
that discovered {\it quarks}, the fundamental constituents of the proton, 
which led to the discovery of Quantum Chromo-Dynamics (QCD) --
the theory of strong interacting color charges
that is responsible for confining colored quarks into all known color neutral hadrons 
by the exchange of {\it gluons}.  
In contrast to the quantum electromagnetism, 
where the force carrying photons are electrically neutral, 
the force carrying gluons carry the color charge 
causing them to interact among themselves, 
which is the defining property of QCD responsible for 
the {\it confinement} of color and the {\it asymptotic freedom} of
color force, two remarkable, while seemingly contradicted, properties of QCD.
It is the HERA at DESY, Germany -- the only electron-proton collider ever built in the world 
that observed the dramatic rise of the probability to find a gluon carrying a small fraction
$x$ of the momentum of colliding proton, which led to a tremendous effort in both 
theory and experiment in recent years to search for a novel phenomenon of gluon saturation
-- an expected consequence of strong gluon fields and the non-linear QCD dynamics. 
Knowing the properties and underlying QCD dynamics of matter at high gluon density 
is critically important for understanding the dynamical origin of 
the newly discovered quark-gluon plasma (QGP) in relativistic heavy ions collisions.
Without gluons there would be no protons, 
no neutrons, and no atomic nuclei.  All matter as we know of would not exist. 
The glue binds us all.  Understanding the properties of the glue and 
its role in determining the internal structure and interactions of nucleons and nuclei,
as well as their emergence from colored quarks and gluons, as dictated by QCD, 
is a fundamental and compelling goal of nuclear science.

In this talk, I briefly review the proposed Electron-Ion Collider (EIC), and 
highlight its science case for building it in the United States.  
Much of the physics presented here are fully documented 
in the EIC White Paper\cite{Accardi:2012qut}, 
put together by a writing committee, 
appointed jointly by the managements of Brookhaven Nationational Laboratory (BNL) 
and Jefferson Laboratory (JLab), with many valuable suggestions and comments from 
the broad nuclear physics community.

%%%%%%%%%%%%%%%%%%%%%%%%%%%%%%%%%%%%%%%%%%%%%%%%%%%%%
% Electron-Ion Collider
\section{Electron Ion Collider}

Since its introduction over forty years ago, QCD has been extremely successful 
in interpreting data from all high energy $e^+e^-$, lepton-hadron and hadron-hadron 
collisions with momentum transfer(s) $\gg \Lambda_{\rm QCD}\sim 1/$fm; and 
{\it ab initio} lattice QCD calculations have provided an impressive description of 
mass spectrum of all known hadrons\cite{Brambilla:2014jmp}.  
While no one questions the validity of QCD today as a fundamental theory of 
strong interactions in the Standard Model, our understanding of QCD remains 
incomplete.  We know very little of the quark-gluon structure of hadrons 
and the emergence of hadrons from quarks and gluons, which are 
closely connected to the defining property of QCD -- the {\it color confinement} 
-- one of the Millennium Prize Problems stated by the Clay Mathematics Institute 
in 2000.

The EIC with its unique capability to collide polarized electrons with polarized protons 
and light ions at unprecedented luminosity, and with heavy nuclei at high energy, 
will be the first precision microscope able to explore how gluons bind quarks to 
form protons and heavier atomic nuclei.  By precisely imaging gluons and sea quarks 
inside the proton and nuclei, the EIC will address some of the deepest and 
the most intellectually pressing questions in QCD:\cite{Accardi:2012qut}
\begin{itemize}
\item {\bf How are the sea quarks and gluons, and their spins,
distributed in space and momentum inside the nucleon?}  How are these
quark and gluon distributions correlated with overall nucleon
properties, such as spin direction?  What is the role of the orbital
motion of sea quarks and gluons in building the nucleon spin?
\item {\bf Where does the saturation of gluon densities set in?}  Is
there a simple boundary that separates this region from that of more
dilute quark-gluon matter? If so, how do the distributions of quarks
and gluons change as one crosses the boundary? Does this saturation
produce matter of universal properties in the nucleon and all nuclei
viewed at nearly the speed of light?
\item {\bf How does the nuclear environment affect the distribution of
quarks and gluons and their interactions in nuclei?}  How does the
transverse spatial distribution of gluons compare to that in the
nucleon? How does nuclear matter respond to a fast moving color charge
passing through it? Is this response different for light and heavy
quarks?
\end{itemize}
Answers to these questions are essential for understanding 
the nature of the visible world.  A full understanding of QCD, 
in a regime relevant to the structure and properties of hadrons and nuclei, 
demands precision measurements at the EIC to explore them in their full complexity. 
Theoretical advances over the past decade have provided quantitative links 
between such measurements and the above questions that 
physicists are trying to answer.  

The EIC will distinguish itself from all past, current, and
contemplated facilities around the world by being at the intensity
frontier with a versatile range of kinematics and beam polarizations,
as well as beam species, allowing the above questions to be tackled at
one facility. The EIC machine designs are aimed at achieving
\begin{itemize}
\item Highly polarized ($\sim$ 70\%) electron and nucleon beams
\item Ion beams from deuteron to the heaviest nuclei (uranium or lead)
\item Variable collision energies, $\sqrt{s}\sim 20\, - \sim$100 GeV,
upgradable to $\sim$140 GeV
\item High collision luminosity $\sim$10$^{33-34}$ cm$^{-2}$s$^{-1}$
\item Possibilities of having more than one interaction region
\end{itemize}
Achieving these challenging technical improvements in a single facility will extend
U.S. leadership in accelerator science and in nuclear science.

%%%%%%%%%%%%%%%%%%%%%%%%%%%%%%%%%%%%%%%%%%%%%%%%%%%%%
% Scientific highlights and golden measurements at the EIC
\section{Scientific highlights and golden measurements at the EIC}
\label{sec:deliverables}

The EIC with its high energy, high luminosity, polarization, and various ion beams
will unite and extend the scientific programs at CEBAF and RHIC in dramatic and 
fundamentally important ways. 

%%%%%%%%%%%%%%%%%%%%%%%%%%%%%%%%%%%%%%%%%%%%%%%%%%%%%
% Nucleon spin
\subsection{Nucleon spin}

\begin{figure}[h!]
\vspace*{-0.4cm}
\begin{center}
\begin{minipage}[b]{0.55\textwidth}
\centering
{\hskip -0.1in}
\includegraphics[width=0.9\textwidth,height=1.75in]{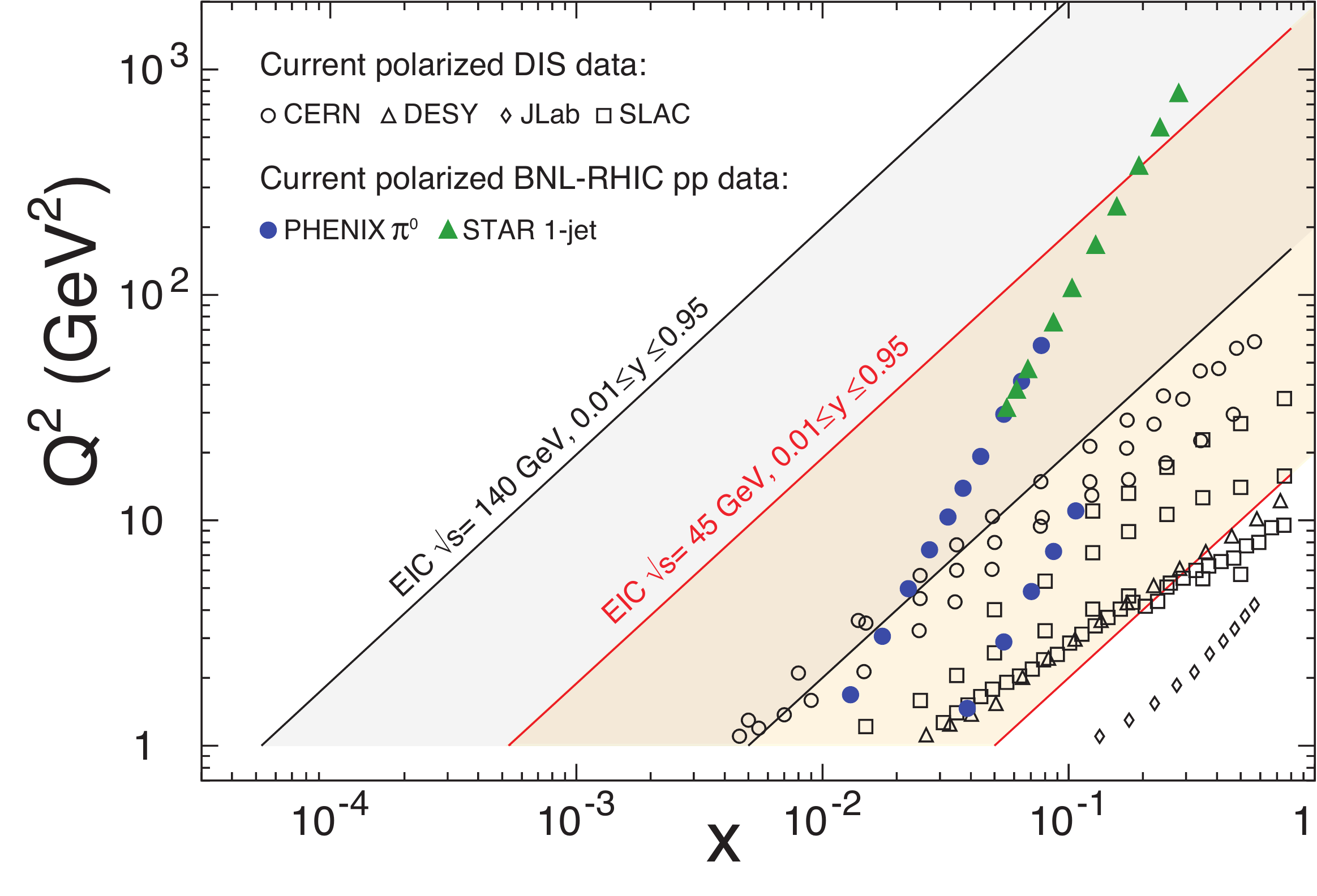}
\end{minipage}
\begin{minipage}[b]{0.44\textwidth}
\centering
\includegraphics[width=0.85\textwidth,height=1.8in]{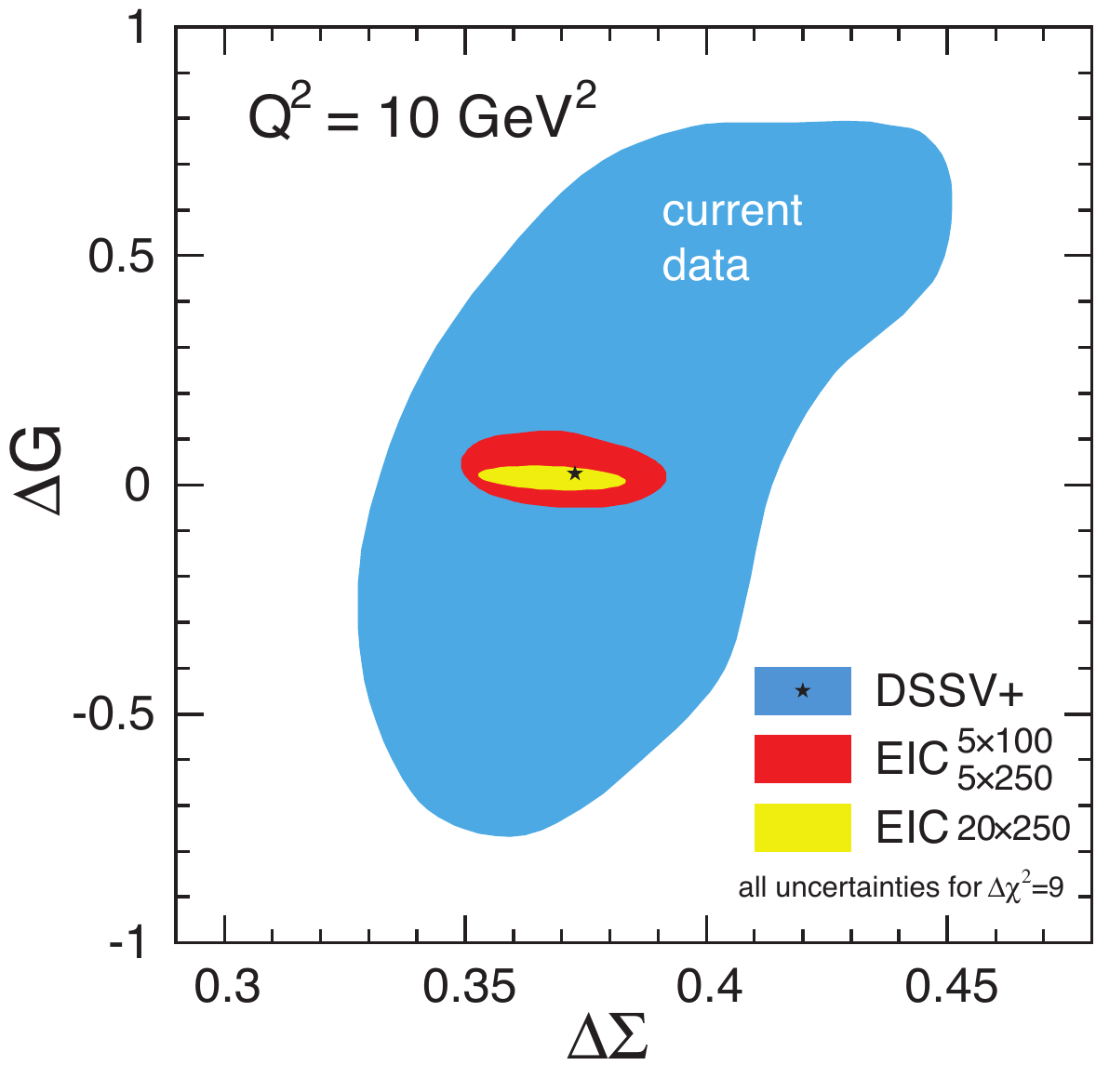}
\end{minipage}
\end{center}
\vskip -0.2in
\caption{\label{fig:spin}{ {\bf Left:} 
The increase coverage in the proton momentum fraction $x$ vs. 
the square of the momentum transferred by the electron to proton $Q^2$ 
at the EIC in $e+p$ collisions. Right: The projected reduction in the uncertainties 
of the gluonÕs helicity ($\Delta G$) and quarkÕs helicity ($\Delta\Sigma$) contributions 
to the protonÕs spin.
}  }
\end{figure}

Several decades of experiments on deep inelastic scattering (DIS) of
electron or muon beams off nucleons have taught us how quarks
and gluons (collectively called partons) share the momentum of a
fast-moving nucleon.  They have not, however, completely resolved the question of
how partons share the nucleon's spin and build up other nucleon
intrinsic properties, such as its mass and magnetic moment. 
Following an intensive and worldwide experimental program over the past two decades,
including recent measurements at RHIC, along with state-of-the-art QCD global analyses,
we learned that the spin of quarks and antiquarks is only responsible
for $\sim 30$\% of the proton spin, while the contribution from the gluons' spin 
is about 20-30\% from the $x$-range explored so far.  
EIC would greatly increase the kinematic coverage in $x$ and $Q^2$, 
as shown in Fig.~\ref{fig:spin}(Left), and hence, reduce the uncertainty of gluonic
contribution to the proton's spin in a very dramatic way, as shown by red and yellow bands
in Fig.~\ref{fig:spin}(Right).  Clearly, the EIC will make a huge impact on our
knowledge of these quantities, unmatched by any other existing or
anticipated facility.  The reduced uncertainties would definitively
resolve the question of whether parton spin preferences alone can
account for the overall proton spin, or whether additional
contributions are needed from the orbital angular momentum of partons
in the nucleon.
%%%%%%%%%%%%%%%%%%%%%%%%%%%%%%%%%%%%%%%%%%%%%
% FIGURE 1
%%%%%%%%%%%%%%%%%%%%%%%%%%%%%%%%%%%%%%%%%%%%%
% END OF FIGURE

\vspace*{-0.6cm}
%%%%%%%%%%%%%%%%%%%%%%%%%%%%%%%%%%%%%%%%%%%%%%%%%%%%%
% Confined motion of quarks and gluons inside the nucleon
\subsection{Confined motion of quarks and gluons inside the nucleon}

\begin{figure}[h!]
\begin{center}
%\vspace{-0.01in}
\begin{minipage}[b]{0.38\textwidth} \centering
{\hskip -0.1in}
\includegraphics[width=0.96\textwidth]{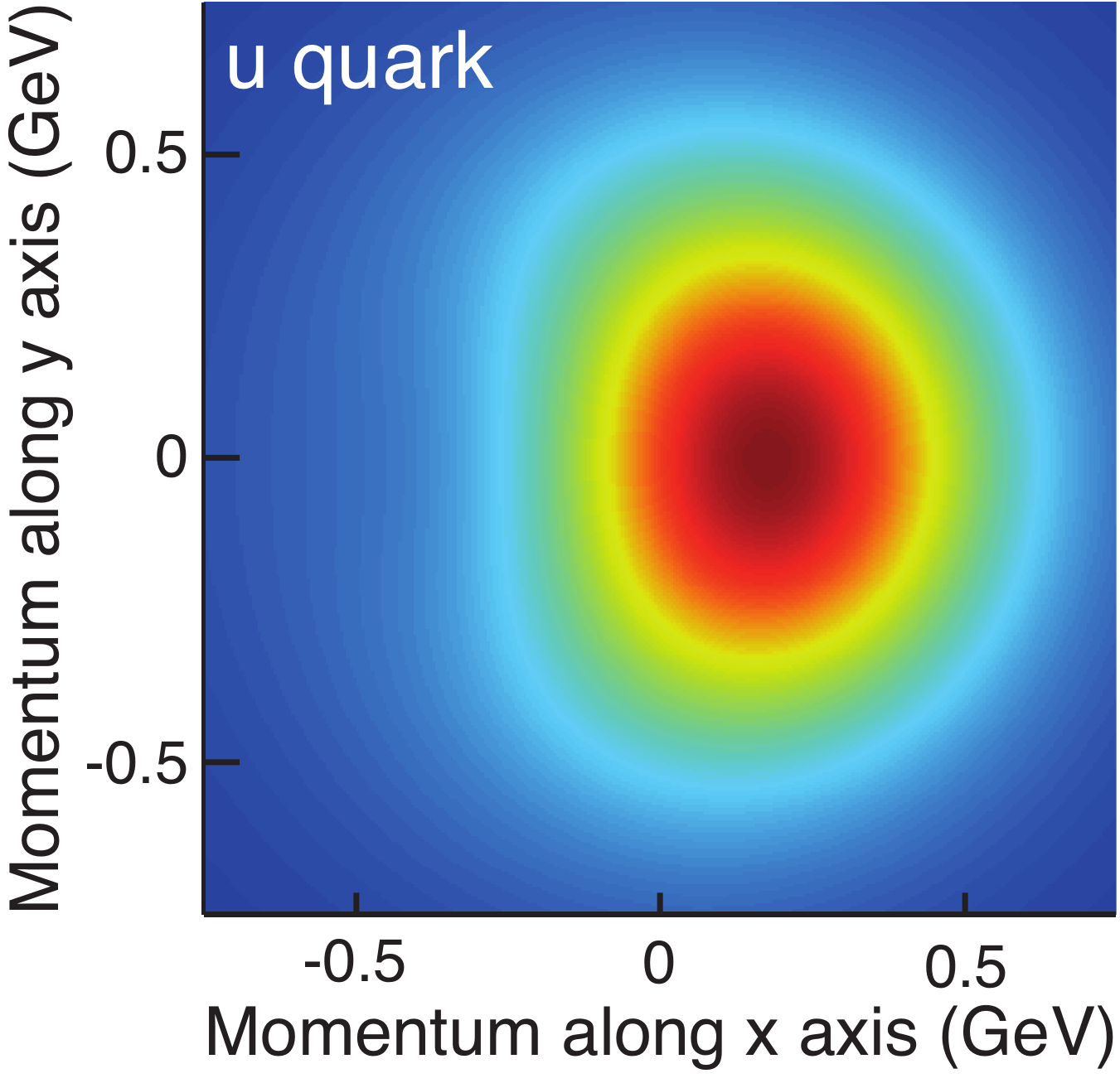}
\end{minipage} 
\hskip 0.05in
\begin{minipage}[b]{0.60\textwidth} \centering
\includegraphics[width=0.98\textwidth]{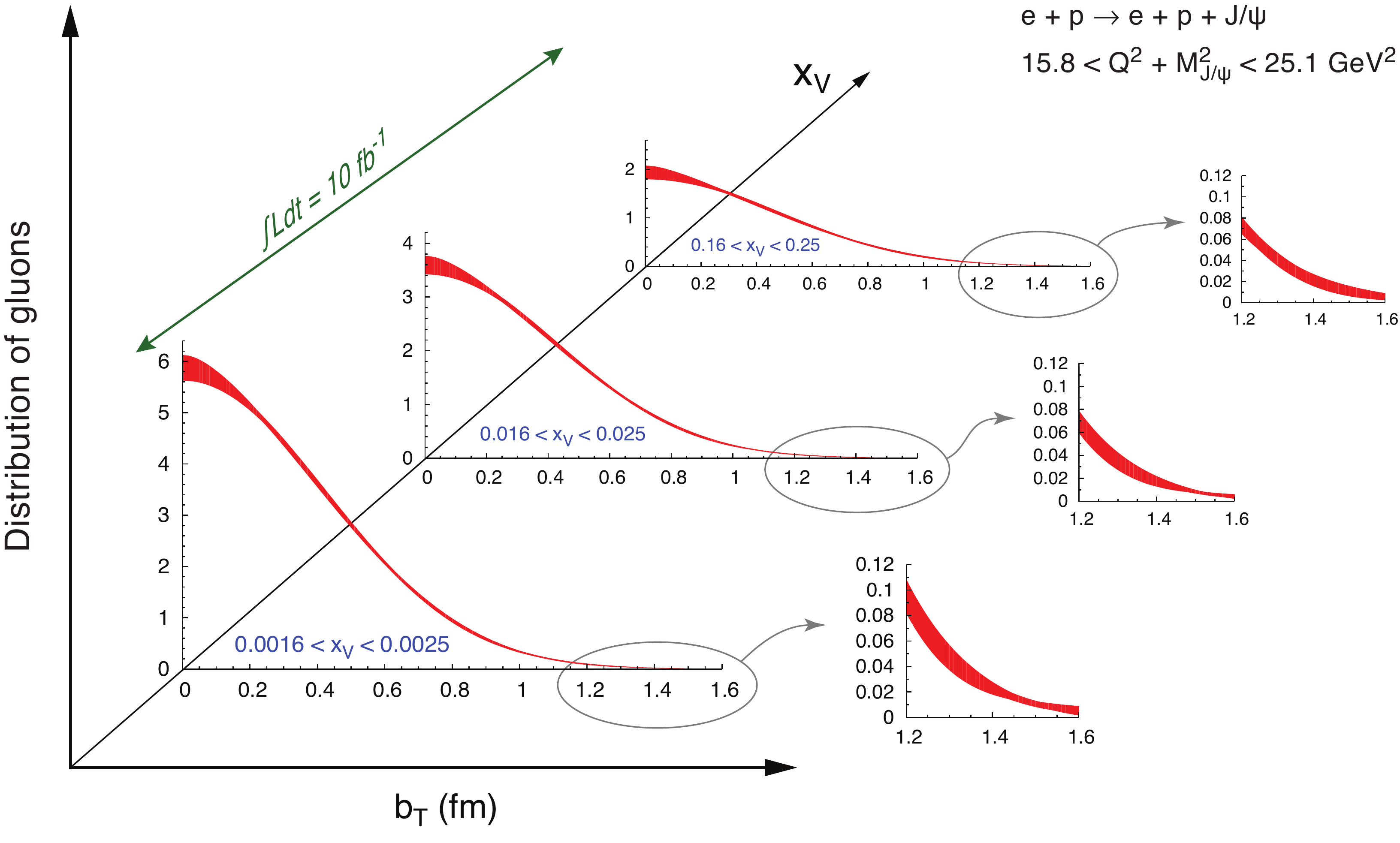}
\end{minipage}
\end{center} 
\vskip -0.05in
\caption{\label{fig:tomography}{ {\bf Left:} The transverse-momentum
distribution of an up quark with longitudinal momentum fraction $x=0.1$
in a transversely polarized proton moving in the $z$-direction, while
polarized in the $y$-direction. The color code indicates the probability
of finding the up quarks.  
{\bf Right:} The projected precision of the transverse
spatial distribution of gluons as obtained from the cross-section of
exclusive $J/\psi$ production at the EIC. }  }
\end{figure}

Semi-inclusive DIS (SIDIS) measurements have two naturally different momentum
scales: the large momentum transfer from the electron beam that ensures
the spatial resolution of the probe, and the momentum of the
produced hadrons perpendicular to the direction of the momentum
transfer that prefers a small value sensitive to the motion of
confined partons.  Remarkable theoretical advances over the past decade
have led to a rigorous framework where information on the confined
motion of the partons inside a fast-moving nucleon is matched to
transverse-momentum dependent parton distributions (TMDs). 
TMDs allow us to investigate the full three-dimensional quark-gluon dynamics 
inside the proton, about which very little is known to date.  
In Fig.~\ref{fig:tomography}(Left), the probability distribution (the color code) to 
find the up quarks inside a proton moving in the $z$-direction
(out of the page) with its spin polarized in the $y$-direction is plotted 
as a function of quark transverse-momentum.  The
anisotropy in transverse momentum is induced by the correlation between the
proton's spin direction and the motion of its quarks and gluons.
While the figure is based on a preliminary extraction of this
distribution from current experimental data, nothing is known about
the spin and momentum correlations of the gluons and sea quarks.  
The EIC will be crucial to initiate and realize such a program, and 
It will dramatically advance our knowledge of the motion of confined gluons
and sea quarks in ways not achievable at any existing or proposed
facility.
%%%%%%%%%%%%%%%%%%%%%%%%%%%%%%%%%%%%%%%%%%%%%
% FIGURE 2
\vspace*{-0.1cm}
%%%%%%%%%%%%%%%%%%%%%%%%%%%%%%%%%%%%%%%%%%%%%
% END OF FIGURE

\vspace*{-0.5cm}
%%%%%%%%%%%%%%%%%%%%%%%%%%%%%%%%%%%%%%%%%%%%%%%%%%%%%
% Spatial Imaging of quarks and gluons inside the nucleon
\subsection{Spatial Imaging of quarks and gluons inside the nucleon}

By choosing particular final states in electron+proton scattering 
without breaking the proton, the EIC with its unprecedented luminosity 
and detector coverage will be able to create detailed spatial images of 
quarks and gluons in the colliding proton. These tomographic images 
obtained from cross-sections and polarization asymmetries for exclusive 
processes are encoded in generalized parton distributions (GPDs), and
are complementary to the one obtained from the TMDs of quarks and gluons, 
revealing aspects of proton structure that are intimately connected with 
the dynamics of QCD at large distances.  
In Fig.~\ref{fig:tomography}(Right), the spatial distribution of gluons, as measured
in the exclusive process: electron + proton $\rightarrow$ electron + proton + $J/\psi$, 
is plotted as a function of $b_T$, the transverse position of the gluons.
Such measurements would reveal aspects of proton structure that 
are intimately connected with QCD dynamics at large distances.

%\vskip 0.1in
%%%%%%%%%%%%%%%%%%%%%%%%%%%%%%%%%%%%%%%%%%%%%
% EA PHYSICS
\subsection{QCD matter at an extreme gluon density}

\begin{figure}[h!]
\begin{center}
%\vspace{-0.01in}
\begin{minipage}[b]{0.49\textwidth} \centering
{\hskip -0.1in}
\includegraphics[width=0.96\textwidth,height=2.1in]{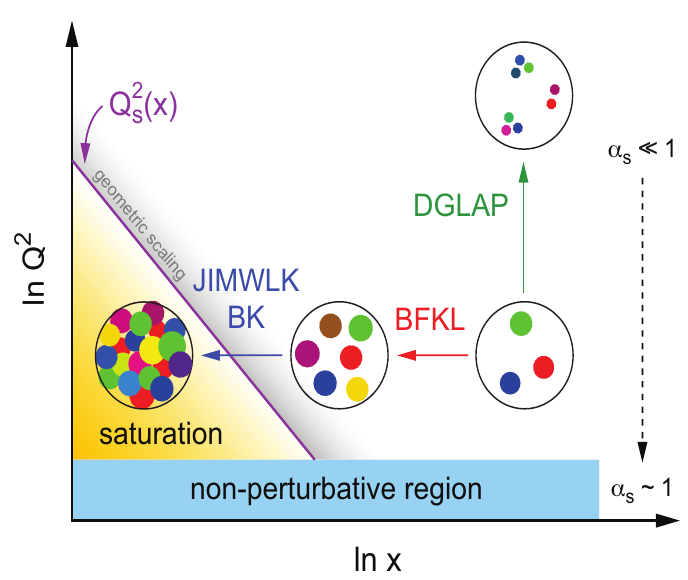}
\end{minipage} 
\hskip 0.05in
\begin{minipage}[b]{0.49\textwidth} \centering
\includegraphics[width=0.98\textwidth,height=2.1in]{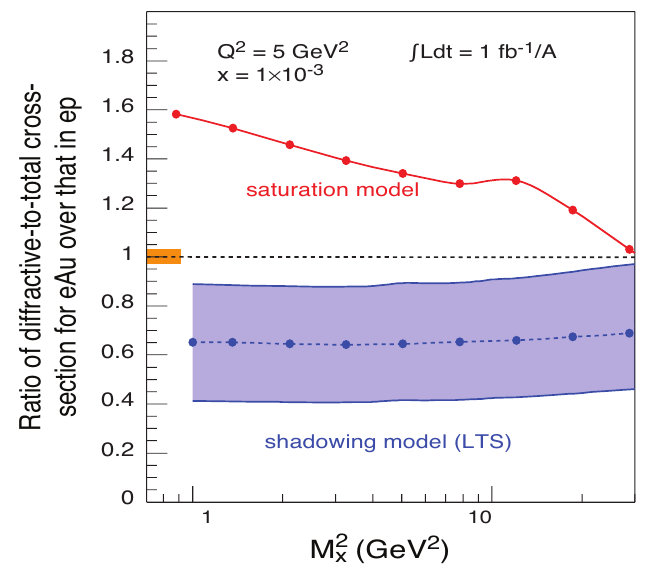}
\end{minipage}
\end{center} 
\vskip -0.1in
\caption{\label{fig:smallx}{ {\bf Left:} The schematic probe resolution vs. energy landscape,
including the regions of low and high saturated parton density, and the
transition region between them.  {\bf Right:} 
The ratio of diffractive
over total cross-section for DIS on gold normalized to DIS on proton
for different values of mass squared of
hadrons produced in the collisions with and without saturation.  
}  }
\end{figure}

When fast moving hadrons are probed by a high resolution local probe, 
the low-momentum gluons contained in their wave functions become experimentally 
accessible. The large soft-gluon density at small-$x$, especially in a large nucleus, 
enables the gluon-gluon recombination to limit the density growth. 
Such a QCD self-regulation mechanism necessarily generates 
a dynamic scale from the interaction of high density massless gluons, 
known as the saturation scale, $Q_s$, at which gluon splitting and 
recombination reach a balance.  At this scale, 
the density of gluons is expected to saturate, producing new and
universal properties of hadronic matter. The saturation scale $Q_s$
separates the condensed and saturated soft gluonic matter from the
dilute, but confined, quarks and gluons in a hadron, as shown in
Fig.~\ref{fig:smallx}(Left). Such universal cold gluon matter 
is an emergent phenomenon of QCD dynamics and of highly scientific interest 
and curiosity.  With its wide kinematic reach, the EIC will be the first experimental facility 
capable of exploring the internal three-dimensional sea quark and gluon structure 
of a fast-moving nucleus, and the transition from a dilute to a dense 
gluon density to provide access to a so far unconfirmed regime of matter 
where abundant gluons dominate its behavior.   
%%%%%%%%%%%%%%%%%%%%%%%%%%%%%%%%%%%%%%%%%%%%%
% FIGURE 3
\vspace*{-0.5cm}
\vspace*{-0.2cm}
%%%%%%%%%%%%%%%%%%%%%%%%%%%%%%%%%%%%%%%%%%%%%
% END OF FIGURE

The existence of such saturated soft gluon matter is a direct consequence
of gluon self-interactions in QCD.  With many more soft gluons at the same impact 
parameters in a large nucleus, enhanced by $A^{1/3}$ with the atomic weight
$A$, than that in a proton, the gluon saturation is much easier to be reached in
$e+A$ than $e+p$ collisions.  By measuring the ratio of diffractive cross-sections 
over the total DIS cross-sections 
in both $e+A$ than $e+p$ collisions, as shown in Fig.~\ref{fig:smallx}(Right), 
the EIC would provide the first unambiguous evidence 
for the novel QCD matter of saturated gluons.  
The EIC is poised to explore with precision the strong gluon field of 
collective dynamics of saturated gluons at high energies. 
Furthermore, knowing the properties of this matter and its underlying QCD dynamics 
is critically important for understanding the dynamical origin of the creation of 
the QGP from colliding two relativistic heavy ions.

%%%%%%%%%%%%%%%%%%%%%%%%%%%%%%%%%%%%%%%%%%%%%
% Shadowing
\subsection{Quark and gluon landscape in a nucleus}

\begin{figure}[h!]
\begin{center}
\begin{minipage}[c]{0.46\textwidth} \centering
\includegraphics[width=0.96\textwidth,height=2.1in]{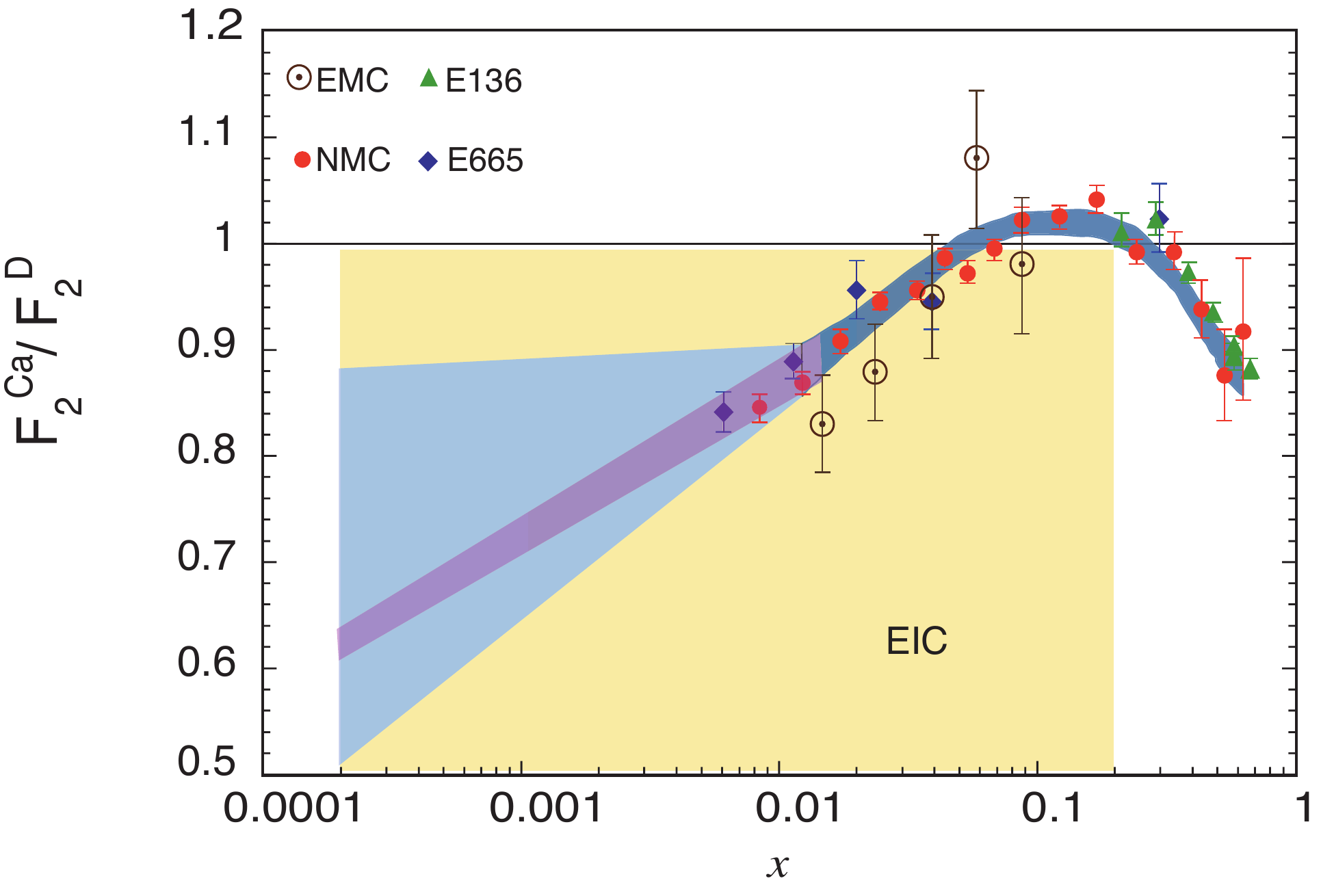}
\end{minipage} \hskip 0.2in
\begin{minipage}[c]{0.46\textwidth} \centering
\includegraphics[width=0.96\textwidth,height=2.0in]{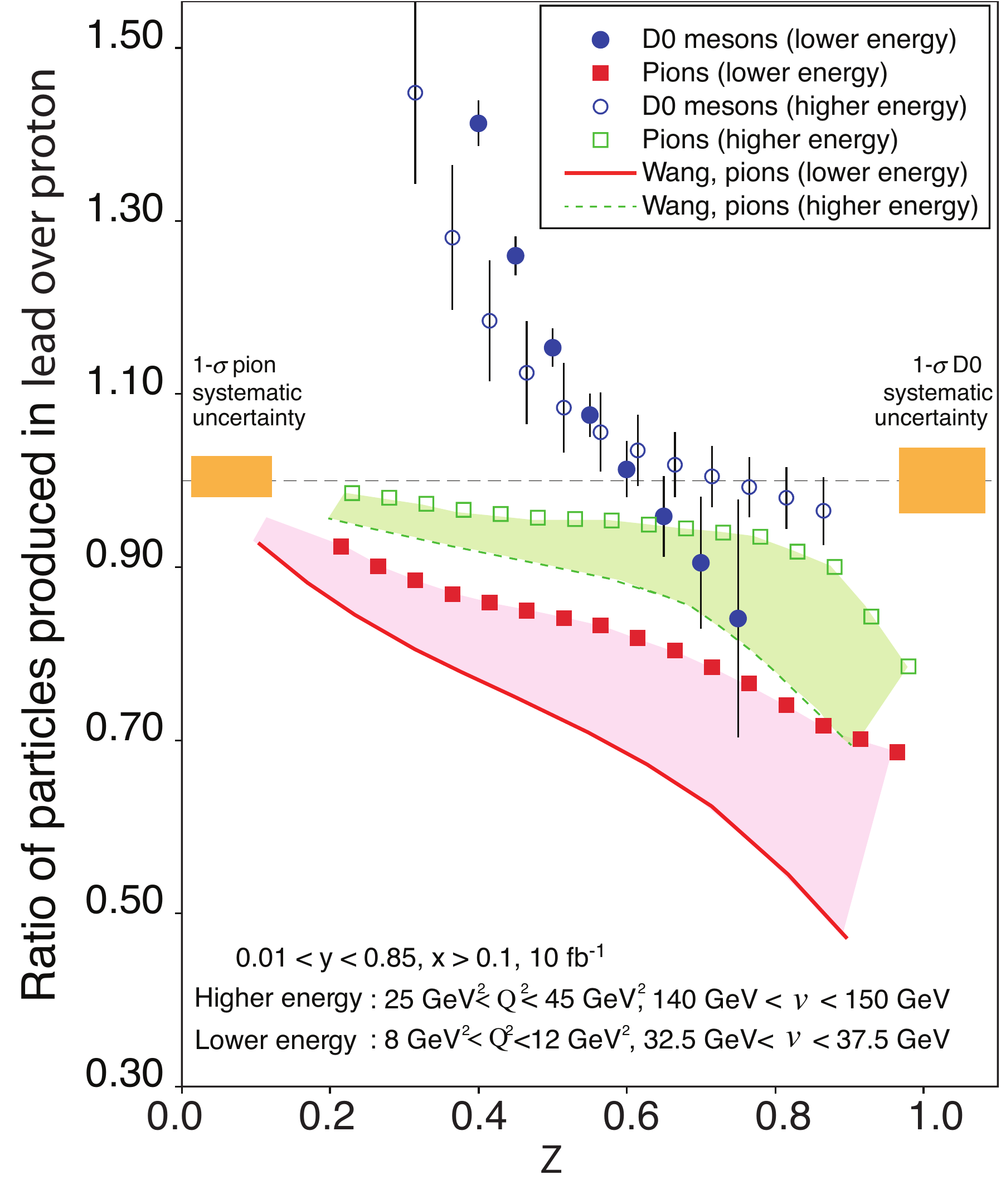}
\end{minipage}
\end{center} \vskip -0.1in
\caption{\label{fig:landscape}{ 
{\bf Left:} The ratio of nuclear over nucleon $F_2$ structure function as a function of
Bjorken $x$, with data from existing fixed target DIS experiments at $Q^2 > 1$~GeV$^2$, 
along with the QCD global fit from EPS09.%\protect\cite{Eskola:2009uj}. 
Also shown is the expected kinematic coverage of the inclusive measurements at the EIC. 
The purple error band is the expected systematic uncertainty
at the EIC assuming a 2\% (a total of 4\%) systematic error, 
while the statistical uncertainty is expected to be much smaller. 
{\bf Right:} The ratio of the semi-inclusive cross-section for producing a pion
(red) composed of light quarks, and a $D^{0}$ meson (blue) composed of
heavy quarks in $e$+lead collisions to $e$+deuteron collisions, plotted as a 
function of $z$, the ratio of the momentum carried by the produced
hadron to that of the exchange virtual photon.  }  }
\end{figure}

The EMC experiment at CERN and experiments in the
following two decades clearly revealed that the distribution of quarks
in a fast-moving nucleus is not a simple superposition of their
distributions within nucleons, as shown in Fig.~\ref{fig:landscape}.  
With its much wider kinematic reach in both $x$ and $Q^2$, the EIC could
measure the suppression of the structure functions to a much lower
value of $x$, approaching the region of gluon saturation.  In
addition, the EIC could for the first time reliably quantify the
nuclear gluon distribution over a wide range of momentum fraction $x$.
%%%%%%%%%%%%%%%%%%%%%%%%%%%%%%%%%%%%%%%%%%%%%
% FIGURE 4
\vspace*{-0.3cm}
%%%%%%%%%%%%%%%%%%%%%%%%%%%%%%%%%%%%%%%%%%%%%
% END OF FIGURE
If the nuclear effect on the DIS cross section, as shown in Fig.~\ref{fig:landscape}, 
is mainly due to the abundance of nucleons at the same impact parameter 
of the nucleus (proportional to $A^{1/3}$), while the elementary scattering is
still relatively weak, one would expect the ratio of nuclear over nucleon structure functions
to saturate when $x$ goes below 0.01, as shown, for example, by the upper
line of the blue area extrapolated from the current data. 
On the other hand, if the soft gluons are a property
of the whole nucleus and the coherence
is strong, one would expect the ratio of the
nuclear to nucleon structure function to fall
continuously as $x$ decreases, as sketched by
the lower line of the blue band, and eventually,
reach a constant when both nuclear
and nucleon structure functions are in the
saturation region. From the size of the purple error band in Fig.~\ref{fig:landscape}, 
which is the expected systematic uncertainty at the EIC 
(while the statistical uncertainty is expected to be much smaller),
the EIC could easily distinguish these two extreme possibilities to
explore the nature of sea quarks and soft gluons in a nuclear environment.

%%%%%%%%%%%%%%%%%%%%%%%%%%%%%%%%%%%%%%%%%%%%%
% Hadronization
\subsection{Hadronization and energy loss}

The exact mechanism of colored partons passing through colored media, 
both cold nuclei and hot matter (the QGP), and the emergence of hadrons 
from the colored partons is not understood. 
A nucleus at the EIC would provide a femtometer filter to 
help determine the correct mechanism by which colored partons 
interact and hadronize in nuclear matter. 
By measuring pion and $D^0$ meson production in 
both $e+p$ and $e+A$ collisions, the EIC would provide 
the first measurement of the quark mass dependence of 
the response of nuclear matter to a fast moving quark.  
The dramatic difference between them, shown in Fig.~\ref{fig:landscape} (Right), 
would be easily discernable at the EIC.  The color bands reflect the limitation 
of our knowledge on hadronization -- the emergence of a pion from a colored quark.  
Enabling all such studies in one place, the EIC will be a true ``QCD Laboratory'', 
unique of its kind in the world.

%%%%%%%%%%%%%%%%%%%%%%%%%%%%%%%%%%%%%%%%%%%%%
% Summary
\section{Summary and outlook}

The EIC is a ultimate machine to discover and explore the three-dimensional 
sea quark and gluon structure of nucleons and nuclei, 
to search for hints and clues of the color confinement, 
to study the color neutralization, and 
to probe the existence of the saturated gluon matter and to explore it in detail. 
The EIC promises to propel both CEBAF and RHIC physics programs to 
the next QCD frontier, and thus, will enable the US to continue
its leadership role in nuclear science research through its quest for
understanding the QCD and gluon-dominated nature of visible matter in
the universe.

%%%%%%%%%%%%%%%%%%%%%%%%%%%%%%%%%%%%%%%%%%
\section*{Acknowledgments}

I thank all members of the EIC White Paper Writing Committee, especially, 
my co-editors, Abhay Deshpande and Zein-Eddine Meziani, for helpful 
discussions and the collaboration.  
This work was supported in part by the U. S. Department of Energy under Contract 
No.~DE-AC02-98CH10886, and the National Science Foundation under 
Grants No.~PHY-0969739 and No.~PHY-1316617.

%%%%%%%%%%%%%%%%%%%%%%%%%%%%%%%%%%%%%%%%%%
%\bibliographystyle{utphysPhysRev}
%\bibliographystyle{h-physrev5}
\bibliographystyle{ws-ijmpcs}
\bibliography{./references/eic_refs}{}
%\begin{thebibliography}{000} %for 3 digits
%\begin{thebibliography}{00}  %for 2 digits
%\begin{thebibliography}{0}    %for 1 digit
%
%\end{thebibliography}

%%%%%%%%%%%%%%%%%%%%%%%%%%%%%%%%%%%%%%%%%%
\end{document}